\documentclass[11pt]{article}
\usepackage[margin=1in]{geometry}
\usepackage{graphicx}
\usepackage[numbers,sort]{natbib}
\usepackage{footnote,enumerate,amsmath,amssymb,amsfonts,amsthm,subcaption,hyperref}

\newcommand*\samethanks[1][\value{footnote}]{\footnotemark[#1]}

\begin{document}
\title{Task-based Parallel Computation of the Density Matrix in Quantum-based Molecular Dynamics using Graph Partitioning\thanks{All authors worked on this project in the Computer, Computational, and Statistical Sciences Division at Los Alamos National Laboratory, Los Alamos, NM 87545, USA, during the 2015 ISTI/ASC Co-Design Summer School.}}
\author{Purnima Ghale\thanks{University of Illinois, Urbana Champaign.}
\and Matthew P.\ Kroonblawd\thanks{University of Missouri, Columbia. Current Address: Physical and Life Sciences Directorate, Lawrence Livermore National Laboratory, Livermore, CA 94550.}
\and Susan M.\ Mniszewski~\thanks{Los Alamos National Laboratory, Los Alamos, NM 87544.}
\and Christian F.A.\ Negre~\samethanks[4]
\and Robert Pavel~\samethanks[4]
\and Sergio Pino~\thanks{University of Delaware, Newark.}
\and Vivek B.\ Sardeshmukh~\thanks{Department of Computer Science, University of Iowa, Iowa City, IA 52242.}
\and Guangjie Shi~\thanks{University of Georgia, Athens.}
\and Georg Hahn~\thanks{Columbia University, New York.}
}
\date{}
\maketitle

\begin{abstract}
Quantum-based molecular dynamics (QMD) is a highly accurate and transferable method for material science simulations. However, the time scales and system sizes accessible to QMD are typically limited to picoseconds and a few hundred atoms. These constraints arise due to expensive self-consistent ground-state electronic structure calculations that can often scale cubically with the number of atoms. Linearly scaling methods depend on computing the density matrix $\mathbf{P}$ from the Hamiltonian matrix $\mathbf{H}$ by exploiting the sparsity in both matrices. The second-order spectral projection (SP2) algorithm is an $O(N)$ algorithm that computes $\mathbf{P}$ with a sequence of $40-50$ matrix-matrix multiplications. In this paper, we present task-based implementations of a recently developed data-parallel graph-based approach to the SP2 algorithm, G-SP2. We represent the density matrix $\mathbf P$ as an undirected graph and use graph partitioning techniques to divide the computation into smaller independent tasks. The partitions thus obtained are generally not of equal size and give rise to undesirable load imbalances in standard \textit{MPI}-based implementations. This load-balancing challenge can be mitigated by dynamically scheduling parallel computations at runtime using task-based programming models. We develop task-based implementations of the data-parallel G-SP2 algorithm using both Intel's Concurrent Collections (\textit{CnC}) as well as the \textit{Charm++} programming model and evaluate these implementations for future use. Scaling and performance results of our implementations are investigated for representative segments of QMD simulations for solvated protein systems containing more than $10,000$ atoms.
\end{abstract}

\section{Introduction}
\label{subsection_outline}
Quantum-based molecular dynamics (QMD) simulations of large physical systems are limited due to the high cost of electronic structure calculations, which often involves determining the eigenstates of the Hamiltonian matrix $\mathbf{H}$. Traditional electronic structure calculations rely on methods such as direct diagonalization, gradient based energy minimization schemes~\cite{genovese2008daubechies}, or Chebyshev filtering techniques~\cite{zhou2006self,banerjee2016chebyshev} that scale as $O(N^3)$, where $N$ denotes the size of the Hamiltonian $\mathbf{H}$. However, instead of individual eigenstates of $\mathbf{H}$, often the density matrix $\mathbf{P}$, which is the sum of the outer product of eigenstates of $\mathbf{H}$, is desired. Elements of $\mathbf{P}$ decay exponentially away from the diagonal for gapped systems and metals at finite temperature, while elements of $\mathbf{P}$ decay algebraically for metallic systems at zero temperature~\cite{goedecker1998decay, ismail1999locality}. The nearsightedness of physical systems, which results in sparse $\mathbf{P}$, makes electronic structure calculations that scale as $O(N)$ possible. An overview of common approaches can be found in the literature, particularly~\cite{bowler2012methods,goedecker1999linear}.

The second-order spectral projection purification method (SP2) is an $O(N)$ algorithm to compute the density matrix $\mathbf{P}$ for any Hamiltonian $\mathbf{H}$. In comparison to other $O(N)$ methods, the advantage of SP2 is that its $O(N)$ scaling is not tied to the method used to generate $\mathbf{H}$. SP2 does not require construction of \textit{Wannier-like} functions with an imposed localization constraint as in~\cite{sena2011linear,haynes2006onetep}, and it also differs from finite-difference based electronic structure calculations that impose localization constraints on numerical solutions~\cite{fattebert2016modeling} to obtain $O(N)$ scaling. While the aforementioned approaches have been shown to be accurate and scaleable, their $O(N)$ scaling is coupled to the choice of basis sets and imposed localization constraints. On the other hand, given any $\mathbf{H}$, SP2 can be used to obtain $\mathbf{P}$ with $O(N)$ computational cost, as long as the matrices are sparse.

SP2 replaces matrix diagonalization by a sequence of matrix-matrix multiplications with approximate $O(N)$~\cite{Mniszewski2015} complexity if a numerically thresholded sparse-matrix algebra is used. Dense SP2 calculations also run efficiently on heterogeneous architectures that include graphics processing units (GPUs) \textit{within a single-node environment}~\cite{CawkwellGPU, cawkwellmultigpu}. However, running the sequential SP2 algorithm on multi-node distributed-memory architectures requires storing, communicating, and updating matrices after every multiplication, and leads to significant overhead even with the help of sparse matrix libraries such as DBCSR~\cite{borvstnik2014sparse}, and SpAMM~\cite{bock2013optimized}. Performance can be improved by noting that if an initial $\mathbf{P}$ is known, the sparse structure of subsequent density matrices remains unchanged over several self-consistency or MD steps. Recently, graph theory was used to re-design the SP2 algorithm so that $\mathbf{H}$ can be divided into smaller matrices such that independent SP2 iterations can be done in a data-parallel way, resulting in substantial performance gains~\cite{niklasson2016graph}.

The graph-based data-parallel version of the SP2 algorithm (hereafter denoted as G-SP2) can significantly improve performance simply by using off-the-shelf graph partitioning methods~\cite{niklasson2016graph}. However, it was found that the subgraphs obtained from off-the-shelf graph partitioning are often of unequal size~\cite{GraphPaper}, which can lead to significant load imbalances in a parallel implementation. These imbalances show wide variability depending on the physical system being studied. For instance, solvated proteins and molecular crystals may have sparse $\mathbf{P}$ but their chemical connectivity varies. If the same QMD application is  to simulate different physical systems efficiently, load balancing should be automated. This becomes a particularly challenging problem in the strong scaling limit, which is needed to reach a low wall-clock time per calculated QMD time step. In this paper, we explore how runtime systems can be used for automatic load-balancing in G-SP2 calculations to fully exploit computing resources. We extend the methods presented in \cite{niklasson2016graph,GraphPaper}, which were designed to minimize the total computational cost of density matrix calculations via G-SP2 but without any considerations about the details of the parallel framework.

\subsection{Parallel implementation of G-SP2}
\begin{figure}[t]
\begin{centering}
\includegraphics[width=0.5\textwidth]{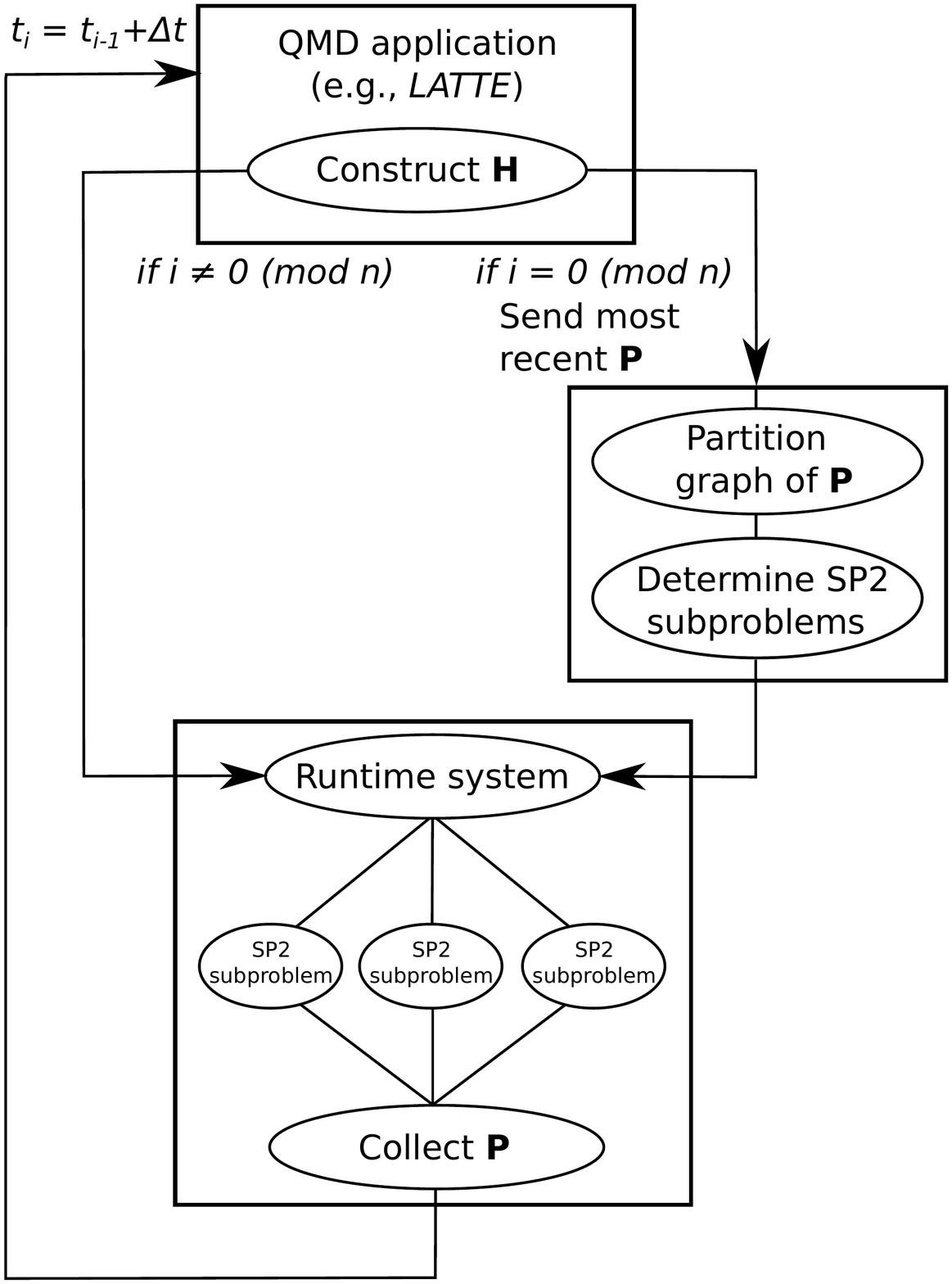}
\caption{Schematic for the proposed data-parallel G-SP2 algorithm in a QMD application such as \textit{LATTE}. The runtime system uses the partitioned graph to divide the SP2 calculation into independent dense matrix-matrix multiplication subproblems. These subproblems are assigned to processors using a task-based scheduler such as \textit{CnC}~\cite{Budimlic2010} or \textit{Charm++}~\cite{Rubensson2014}. The resulting $\mathbf{P}$ is collected and sent back to the QMD application.}
\label{fig:schematic}
\end{centering}
\end{figure}

Fig.~\ref{fig:schematic} shows a schematic of our proposed parallel implementation of the G-SP2 algorithm using a task-based runtime system. Initially, an external QMD method, such as the self-consistent density functional tight-binding scheme as implemented in \textit{LATTE}~\cite{LATTE}, computes the self-consistent Hamiltonian matrix $\mathbf{H}$ and its corresponding density matrix $\mathbf{P}$. This is done using the standard numerically thresholded sparse SP2 scheme, which is hard to parallelize over many nodes. The non-zero structure of the thresholded density matrix $\mathbf{P}$ is then used to determine the adjacency matrix of a graph that approximates the data dependencies in the Fermi-operator expansion. Partitioning algorithms are used to divide the graph into subgraphs that are used to extract submatrices of $\mathbf{H}$ corresponding to independent small dense subproblems. The subproblems identified through graph partitioning serve as inputs to a runtime system that dynamically schedules dense matrix SP2 computations that run independently until convergence. The resulting submatrices are collected and reassembled to form the full density matrix $\mathbf{P}$, which is subsequently used to estimate the data dependency graph and partitioning for the next time step, possibly adding or removing data dependencies as the atoms have moved.

Simulated material systems do not exhibit rapid changes on the order of the integration step size, meaning that the graph structure extracted from $\mathbf{P}$ is largely preserved over several time steps. It is therefore reasonable to reuse a particular set of graph partitions over a user-specified number of time steps $n$ chosen with respect to the system kinetics and dynamics. This is analogous to periodically updated neighbor lists for computing pairwise interactions in standard MD implementations~\cite{FrenkelSmit}. The variable $i$ in Fig.~\ref{fig:schematic} serves as an iteration counter: in each iteration $i$, a new $\mathbf{P}$ is computed at time $t_i$ in a parallel fashion. However, the partitions are only re-computed every $n$ steps (i.e., the branch $i=0$ (mod) $n$ is executed). Otherwise, $\mathbf{P}$ is computed according to the existing (old) partitions corresponding to the branch $i \neq 0$ (mod) $n$.

\section{\label{section:motivation} Methods}
\subsection{Second-Order Spectral Projection Method (SP2)}
The SP2 algorithm~\cite{niklasson2004density} computes $\mathbf{P}$ via matrix multiplications without explicitly computing the chemical potential $\mu$ or the eigenstates of $\mathbf{H}$. The SP2 algorithm starts from an initial guess for the density matrix, $\mathbf{X}_0$, which is a rescaled Hamiltonian 
\begin{equation}
\mathbf{X}_0 = \frac{\epsilon_{\text{max}} \mathbf{I} - \mathbf{H}}{\epsilon_{\text{max}} - \epsilon_{\text{min}}},
\end{equation}
whose eigenvalues are mapped to the interval $[0,1]$ in reverse order. Estimates of the minimum ($\epsilon_{\text{min}}$) and maximum ($\epsilon_{\text{max}}$) eigenvalues of $\mathbf{H}$ were obtained from the Gershgorin circle theorem~\cite{Gerschgorin1931}. Gershgorin circles can be used here because only the maximum and minimum bounds, and not their accurate values are required. At each iteration $i$, an $\mathbf{X}_i$ is obtained from the previous estimate $\mathbf{X}_{i-1}$ using
\begin{equation}
\mathbf{X}_i = \left[ \mathbf{I}+ \alpha_i (\mathbf{I} - \mathbf{X}_{i-1}) \right] \mathbf{X}_{i-1}.
\label{eq:Iteration}
\end{equation}
Here, $\alpha_i = 1$ if $2\text{Tr}[\mathbf{X}_i] \leq N_{\text{el}}$, where $N_\text{el}$ denotes the number of electrons of the physical system, and $\alpha_i = -1$ otherwise. In self-consistent charge tight-binding, the initial density matrix is computed using regular SP2 from which we store the multiplication sequence determined by $\{ \alpha_i \}_{i=1,2,\ldots}$. This sequence of multiplications can be used to compute the density matrix during the iterative refinement of $\mathbf{H}$ and $\mathbf{P}$. The same sequence is used either until convergence of the self-consistent charge procedure or for a fixed number of iterations. In case the sequence of $\alpha_i$ is wrong, the iteration ends with the wrong trace, and the sequence has to be recomputed. $\mathbf{P}$ is obtained in the limit
\begin{equation}
\mathbf{P} = \lim_{i\rightarrow \infty} \mathbf{X}_i.
\label{sp2limit}
\end{equation}
A criterion to check the convergence in eq.~(\ref{sp2limit}) is to verify that $\mathbf{P}$ is idempotent (i.e., $\mathbf{P}^2 = \mathbf{P}$) within given error bounds~\cite{SP2a}. In our examples, a few dozen iterations were usually sufficient to compute $\mathbf{P}$, even for systems as large as $\sim10,000$ atoms.

\subsection{Graph Theory}
\label{GPmethods}
After the initial density matrix $\mathbf{P}$ has been computed using regular SP2, a graph $G_{\tau}=(V,E)$ with vertex set $V$ and edge set $E$ is constructed by using the non-zero structure of $\mathbf{P}$ thresholded with a global parameter $\tau$~\cite{niklasson2016graph}. Each row (or column, since $\mathbf{P}$ is symmetric) of $\mathbf{P}$ corresponds to a vertex $v$ in the graph. Pairs of vertices $u$ and $v$ will be connected through an edge $(u,v)$ in $E$ if $|\mathbf{P}_{u\,v}|>\tau$. The threshold $\tau$ is a parameter used in both graph-based and standard~\cite{cawkwell2012energy} SP2 implementations which allows for a well-controlled tuning of the computational error.

\begin{figure*}
  \begin{minipage}{0.49\textwidth}
    \centering
    \includegraphics[width=0.5\textwidth]{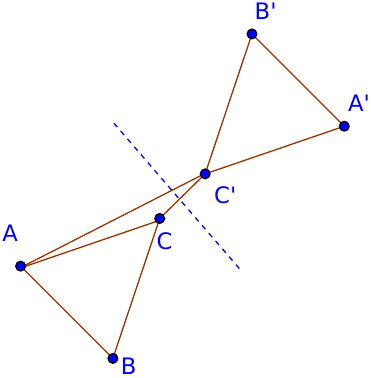}
    \subcaption{\label{fig:corehaloa} Each partition has three core vertices and one neighbor vertex in the other partition.}
  \end{minipage}\hfill
  \begin{minipage}{0.49\textwidth}
    \centering
    \includegraphics[width=0.5\textwidth]{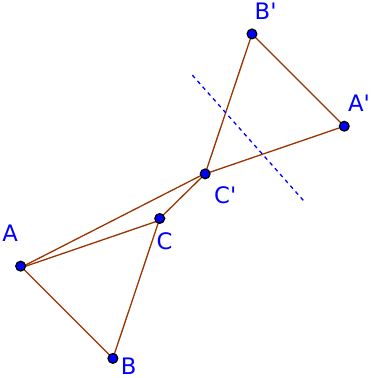}
    \subcaption{\label{fig:corehalob} One partition has 4 core vertices and two neighbors. Another partition has two cores and one neighbor.}
  \end{minipage}
\caption{Two ways to partition the same graph resulting in equal edge-cuts but with different subproblem sizes. The computational cost of G-SP2 is proportional to the cubed sum of the number of vertices and neighbors in each partition. The partitioning shown in (a) is thus preferred.} 
\label{fig:corehalo}
\end{figure*}

Graph partitioning algorithms divide a graph into distinct partitions while attempting to minimize a variety of metrics. Let $G$ be an undirected graph and $n \geq 2$ be an integer denoting the number of partitions (or equivalently, subproblems). We call any vertex belonging to a given graph partition a \textit{core vertex} of that partition. Additionally, we are interested in the set of all vertices that do not belong to the partition, but are connected neighbors of any of the core vertices. We will call any such vertex a \textit{halo vertex} of the partition. Each G-SP2 subproblem $i$ is comprised of $c_i$ core vertices and $h_i$ halo vertices in neighboring partitions, so its size is $s_i = c_i + h_i$. We are interested in partitioning $G$ into $n$ partitions in a way that minimizes the objective function
\begin{equation}
\label{eq:sumOfCubes}
C( \{c_i\}, \{h_i\}; n) = \sum_{i=1}^n (c_i+h_i)^3,
\end{equation}
which is the total computational cost for the G-SP2 subproblems. (Or more precisely, the total cost of the corresponding dense matrix-matrix multiplications for each submatrix of $\mathbf{P}$ corresponding to a partition.) Partitioning a graph with respect to the objective function in eq.~\eqref{eq:sumOfCubes} was called \textit{CH-partitioning} (core-halo partitioning) in \cite{GraphPaper}. Note that this objective function differs from the one employed in traditional edge-cut minimization \cite{KarypisKumar1999,sandersschulz2013}. We observe this by investigating two partitionings (each with two partitions) of a simple graph in Fig.~\ref{fig:corehalo}.

In the first scenario (Fig.~\ref{fig:corehaloa}), the two partitions consist of the vertices $(A,B,C)$ and $(A',B',C')$, respectively. In the second scenario (Fig.~\ref{fig:corehalob}), the graph is divided into partitions containing $(A,B,C,C')$ and $(A',B')$. Edge-cut minimization cannot distinguish between the cases depicted in Figs.~\ref{fig:corehaloa} and \ref{fig:corehalob} because the number of edges between the two partitions is two in both cases. 

In Fig.~\ref{fig:corehaloa}, one of the two G-SP2 subproblem has core vertices $A$, $B$, and $C$ and halo vertex $C'$ while the other has core vertices $A'$, $B'$, and $C'$ and halo vertices $A$ and $C$. This results in two submatrices of dimension $4$ and $5$ respectively. In Fig.~\ref{fig:corehalob}, one subproblem contains four core vertices and two halo vertices, whereas the other subproblem contains two core and one halo vertex, resulting in two submatrices of dimension $6$ and $3$, respectively. The computational cost for G-SP2 using Fig.~\ref{fig:corehaloa} is proportional to $4^3 + 5^3 = 189$, while the computational cost for Fig.~\ref{fig:corehalob} is proportional to $6^3 + 3^3 = 243$.

In \cite{GraphPaper}, the standard graph partitioning software packages \textit{METIS} \cite{KarypisKumar1999}, \textit{KaHIP} \cite{sandersschulz2013} and \textit{hMETIS} \cite{KarypisKumar2000} were evaluated with respect to their performance in computing efficient CH-partitionings. In addition, an approach based on simulated annealing (SA) designed to minimize eq.~\eqref{eq:sumOfCubes} was also evaluated. From \cite{GraphPaper}, we conclude that in practice, \textit{METIS} was most suitable for efficiently minimizing eq.~\eqref{eq:sumOfCubes} and thus for computing CH-partitionings on a variety of test graphs. Simulated annealing almost always improved the quality of partitions, and was more useful for dense graphs. However, using SA could lead to more load imbalances, as entire partitions were 'dissolved' in some cases in order to improve the numerical cost metric. Thus, METIS was used to partition graphs for all the results presented in this manuscript.

Finally, we note that independent subproblems can also be constructed by using locality -- core vertices in partitions can be chosen based on locality, and given these core vertices, there will be a localization radius beyond which elements of $\mathbf{P}$ can be set to zero as in \cite{fattebert2016modeling} -- but graph partitioning provides a more general way to obtain partitions (e.g., by partitioning using an objective function which can, but does not need to, depend on distances). Grouping matrix elements according to their interaction strengths in $\mathbf{P}$ rather than the distance can be especially useful for sparse and heterogeneous simulations cells, where distance-based cutoffs may result in larger submatrices than necessary. 

\subsection{Asynchronous Programming Models}
G-SP2 allows us to divide a large sequential problem into smaller data-parallel subproblems that can be scheduled using a Bulk Synchronous Parallel (BSP) implementation: for instance, by using \textit{MPI} (message passing interface) and \textit{OpenMP} threading. This approach is sufficient for uniform systems, where subproblems can be expected to be of similar sizes, or when a specific physical system is being targeted so that possible load imbalances are known a priori and can be mitigated. However, for a general application used to study systems with a variety of bonding configurations a greater variability in subproblem sizes is expected, and it is for these systems that asynchronous parallel programming based G-SP2 will be most useful. The type and severity of load imbalance varies with the physical system being studied; this rules out hard-coding a scheduler or load balancer.
\begin{figure}
\begin{center}
\includegraphics[width=0.5\columnwidth]{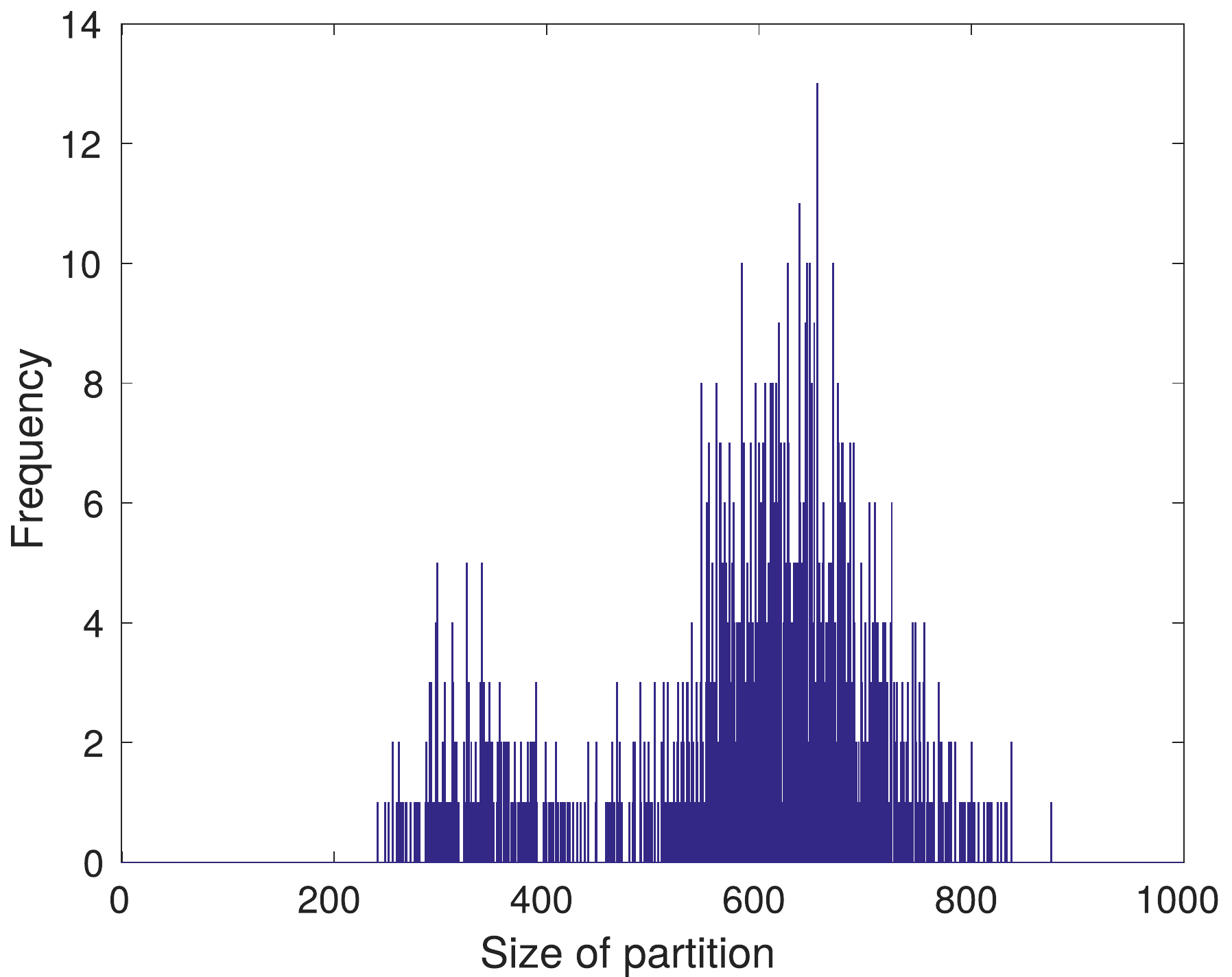}
 \caption{\label{fig:subgraph-size} Distribution of G-SP2 subproblem matrix sizes obtained using simple block partitioning applied to the density matrix of \textit{polyalanine 289} having dimension $\sim40,000$.}
 \end{center}
\end{figure}

Fig.~\ref{fig:subgraph-size} shows an example for the distribution of G-SP2 subproblem sizes where load-balancing can be useful. The density matrix of \textit{polyalanine 289} with $41185$ vertices, one of the test matrices used in \cite{GraphPaper}, was block partitioned into 1200 partitions. By block partitioning, we refer to a simple technique which allocates a fixed number of cores (or orbitals) to each partition in order of their appearance in the Hamiltonian matrix. For the polyalanine molecule, each partition either has $34$ or $35$ cores. Although each partition contains about the same number of core vertices, the number of halo vertices for each partition differs significantly and affects the subproblem size. In more densely connected graphs, each core vertex has more neighbors so that given an equal number of core vertices per partition, subproblems are likely to be larger, and the distribution in Fig.~\ref{fig:subgraph-size} shifts to the right. Similarly, for heterogenous systems with the same core vertices in each partition, but with coexisting sparsely and densely connected regions, a wider distribution of subproblem sizes is expected.

\section{Results}
\subsection{Selection of a runtime system}
Asynchronous parallel programming is relatively new to scientific applications so we investigated two asynchronous task-based runtime systems, namely Intel Concurrent Collections \textit{CnC} \cite{Budimlic2010} and \textit{Charm++} \cite{Rubensson2014}, to schedule and execute SP2 subproblems. Detailed descriptions of \textit{CnC} and \textit{Charm++} are found in Appendix~\ref{long_task}. The Los Alamos National Laboratory CentOS cluster Darwin, which has a highly heterogeneous architecture with four 18-core Intel Xeon E7-8880 v3 CPUs clocked at 2.3~GHz and 512~GB of RAM, was used. Hyper-threading was disabled so that each node had 72 cores and 72 threads. All runtimes in this and the following sections are for a single density matrix computation. 

\begin{figure}[ht!]
\begin{center}
\includegraphics[width=0.7\columnwidth]{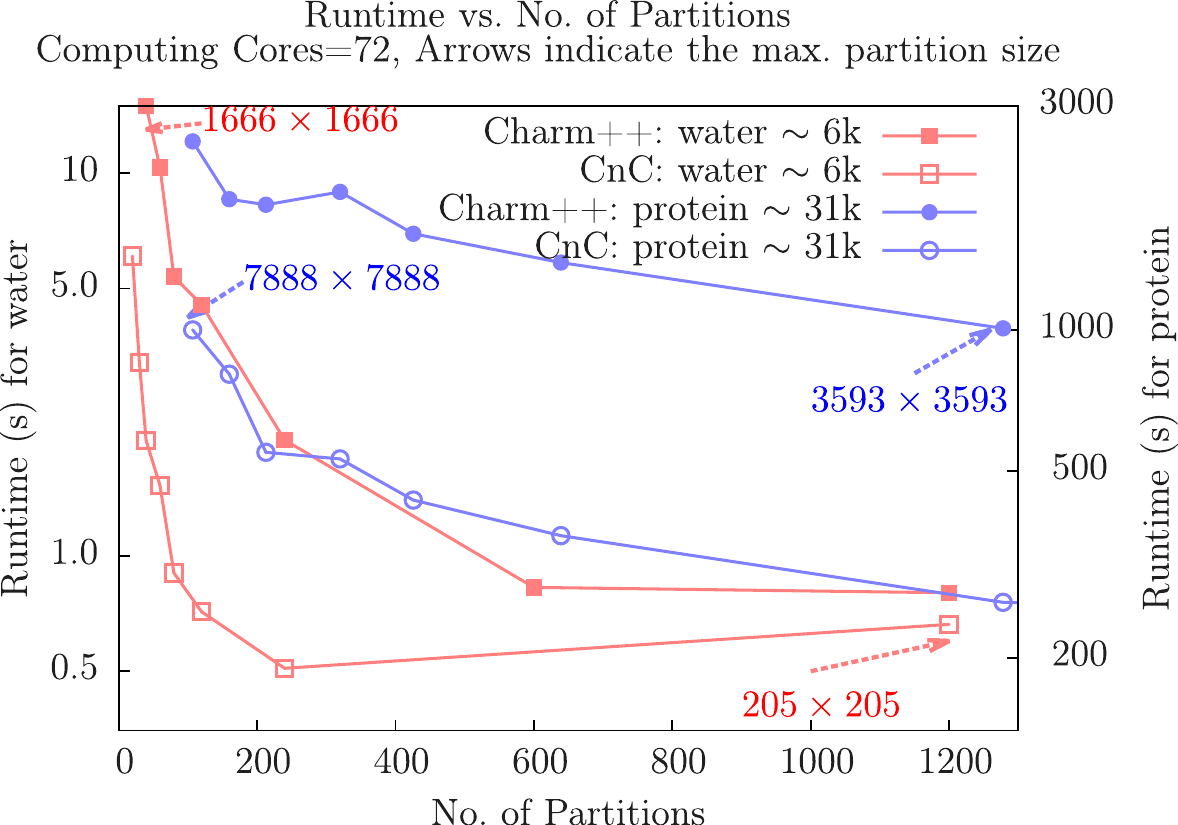} 
\caption{Runtime for density matrix computation as a function of number of partitions for \textit{CnC} and \textit{Charm++} implementations applied to a liquid water system (runtime scale on left y-axis) with $\mathbf{P}$ size $\sim6000$ and a solvated protein system (runtime scale on right y-axis) with $\mathbf{P}$ size $\sim 31,000$. The largest subproblem size for selected data points is indicated with arrows.}
\label{fig:time-part}
\end{center}
\end{figure}

Fig.~\ref{fig:time-part} shows the time taken for a single density matrix computation as a function of the number of partitions for two different systems. Runtimes on the left y-axis correspond to a 3D-periodic simulation cell containing liquid water with density matrix of size $\sim6000$, while runtimes on the right y-axis correspond to a unit cell of protein solvated in water with $\mathbf{P}$ of size $\sim 31,000$. Arrows within Fig.~\ref{fig:time-part} indicate the largest subproblem sizes, which represent the most expensive linear algebra computations, for the specified numbers of partitions and for both simulation cells. We note that partitioning the SP2 calculation into subproblems first results in a significant decrease in runtime but as the number of partitions is increased, the matrix can be over-decomposed resulting in worse performance as reflected in the \textit{CnC} simulation for the smaller water system. 

Fig.~\ref{fig:time-part} also shows that our \textit{CnC} application is significantly more efficient than the \textit{Charm++} one in a shared-memory environment. \textit{Charm++} was developed for distributed memory using charm-MPI interoperability, while the \textit{CnC} version is for shared memory, and requires significant code changes to be able to run with distributed nodes. Indeed, in experiments the \textit{Charm++} application spawned N processes, while the \textit{CnC} application spawned N threads. As a result, the \textit{Charm++} application spent more time on communication and overhead as reflected in its runtime. However, while \textit{CnC} outperformed \textit{Charm++} in a shared memory environment, it could not be ported easily to run over distributed nodes. The ability to run on distributed nodes is essential if large systems that do not fit on a single node are to be simulated in the future, and this extension to distributed nodes was straightforward with \textit{Charm++}. Therefore, we focus our efforts on using optimized subproblems for G-SP2 computed with \textit{Charm++} while noting that ideally, single-node performance would also be optimized.

\begin{figure}[t]
  \begin{subfigure}{0.49\linewidth}
  \includegraphics[width=\textwidth]{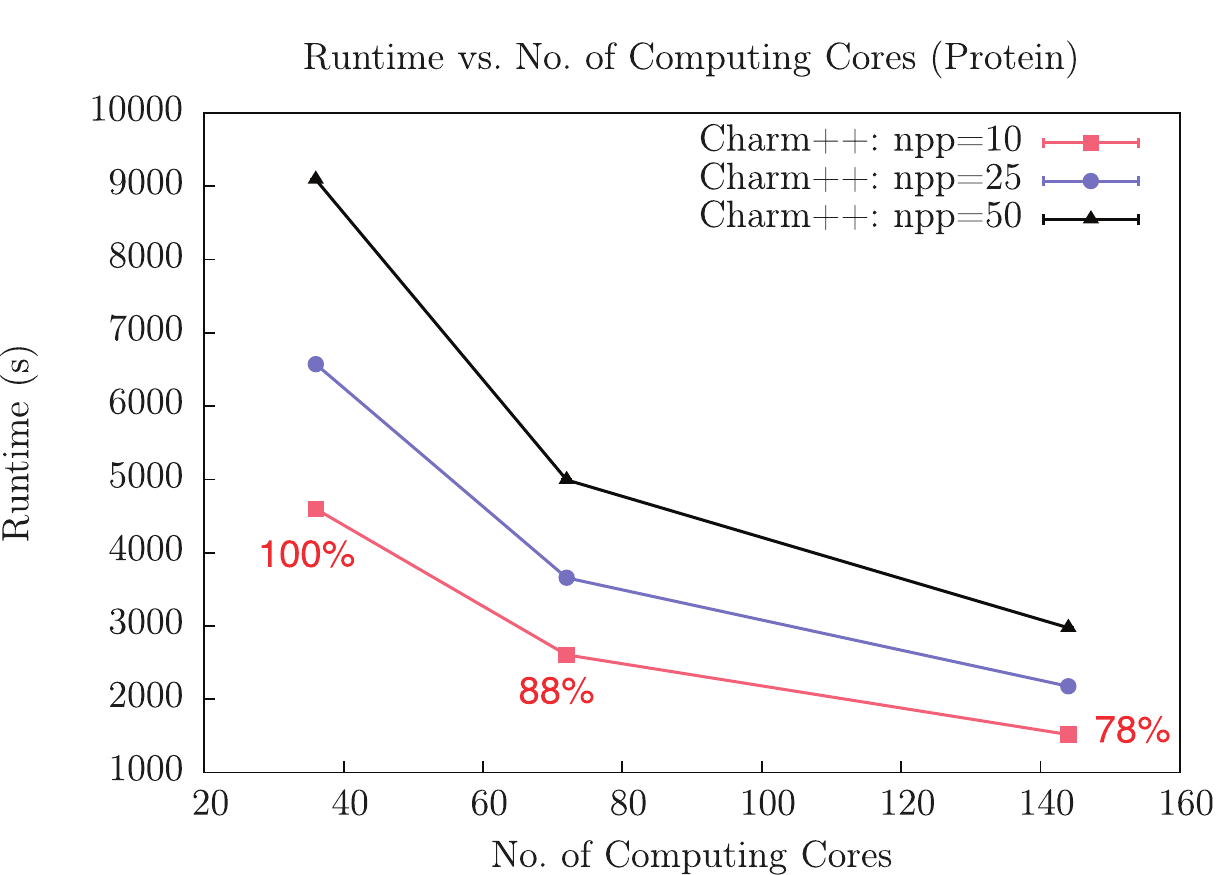}
  \caption{\label{fig:time-core-2}}
  \end{subfigure}~
  \begin{subfigure}{0.49\linewidth}
  \includegraphics[width=\textwidth]{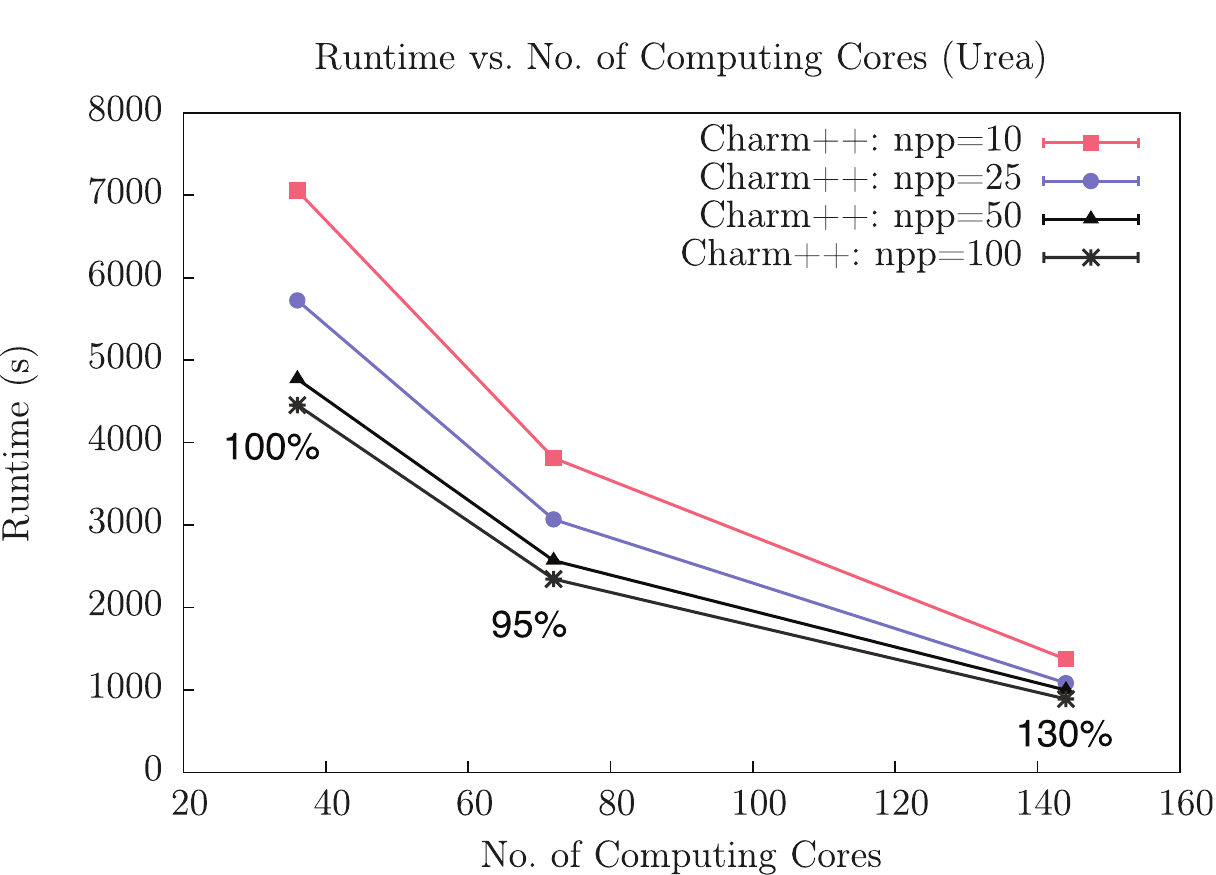}
 \caption{\label{fig:time-core-3}}
  \end{subfigure} 
  \caption{Runtimes of our \textit{Charm++} implementation for (a) a solvated protein (sparse) and (b) the molecular crystal urea (dense). As the number of partitions per processor is increased from 10 to 50 partitions per processor, runtimes for the protein system (a) increase, indicating over-decomposition beyond the optimal number of partitions, while runtimes for the more densely connected urea system (b) decrease. Parallel efficiency for the number of partitions with lowest runtimes are indicated; superlinear scaling is observed in the case of the urea simulation with 144 cores.\label{fig:time-core}}
\end{figure}

Fig.~\ref{fig:time-core} shows the performance of our \textit{Charm++} implementation with block-partitioned subgraphs on distributed nodes for two systems with $\mathbf{P}$ of similar size: a solvated protein system (Fig.~\ref{fig:time-core-2}), and a molecular crystal urea (Fig.~\ref{fig:time-core-3}) that was previously considered in \cite{GraphPaper} and was chosen here because it has a more densely connected $\mathbf{P}$ compared to the solvated protein.

We observe that as the number of partitions per processor is varied (10, 25 and 50 partitions per processor), the runtime for the solvated protein in Fig.~\ref{fig:time-core-2} increases, while the runtime for the crystal urea system decreases in Fig.~\ref{fig:time-core-3}. Given the same number of block-partitions, dense systems have more neighbors than sparse ones, so that denser systems lead to larger submatrices once neighbors are taken into account as halos. The size of submatrices determines data re-use (which mitigates communication costs), while the linear-algebraic cost scales cubically with submatrix size. Therefore, the optimal number of partitions balances communication and computational costs, and will vary with the system. In particular, comparing Figs.~\ref{fig:time-core-2} and \ref{fig:time-core-3}, we observe that with 10 partitions per processor, runtimes for the dense crystal urea system are considerably higher than for the sparse solvated protein system. Assuming communication costs have not dominated, this behavior is expected, as denser systems give rise to larger submatrices and higher computational costs for dense matrix multiplication. However, as the number of partitions is increased, submatrices of both physical systems get smaller; the sparse system is soon over-decomposed into more than the optimal number of partitions, while performance improves for the dense system, as there are still gains to be made by increasing the number of partitions.

Finally, the parallel efficiency was computed using \cite{xavier1998introduction}: \[\text{parallel efficiency} = \frac{\text{speedup }}{\text{ratio of processors used}} =  \frac{ \left( \frac{t_b}{t_n} \right) }{ \left( \frac{n}{b} \right) }\times100 \%,\] where $t_n$ is the runtime for simulations using $n$ computing cores and $b=36$ is a baseline number against which the performance is compared (our implementation was tested for a minimum of 36 computing cores). Although we were unable to reach the strong scaling limit due to the number of computing cores available at the time, superlinear scaling is observed in the case of the crystal urea simulation with 144 cores. Superlinear scaling is uncommon but has been observed occasionally in applications~\cite{lange2013achieving,hess2008gromacs}. In particular, this behavior is to be expected when most or all of the working set (i.e., the amount of memory required for the process) can fit on the total available cache, and it is a known cause of superlinear scaling~\cite{gustafson1990fixed}. As the number of cores is increased, the parallel efficiency is expected to again fall below $100\%$.

\subsection{Evaluation of G-SP2 with graph partitioning and the \textit{Charm++} runtime system}
\label{section_fullqmd}

The G-SP2 algorithm in connection with both the \text{blocking scheme} and the \textit{METIS} graph partitioning package were implemented using \textit{Charm++}. SP2 computations were performed for the \textit{polyalanine 289} protein system previously studied in \cite{GraphPaper} as well as in previous sections of this paper. The computing cluster used for these computations was a Dell-built cluster running Linux with $14$ \textit{R730} computing nodes. All $14$ nodes were equipped with two $12$-core Intel Haswell CPUs, 64~GB of RAM ($2 2/3$ GB/core), a 200~GB solid state drive, gigabit ethernet, dual port FDR infiniband, and access to a shared $100+$ TB RAID storage system. Our implementation was built using \textit{Charm++} version $6.7.0$ and the \textit{RefineCommLB} load balancer. This choice of the load balancer has been shown to yield robust performances on various inputs \cite{Pilla2015} and outperformed another popular choice, \textit{GreedyCommLB}, in our experiments. Dense matrix-matrix multiplications were performed using the \textit{dgemm} routine from \textit{mkl cblas} (version $16.0.0$). All performance comparisons in this section are described in terms of the average overall runtime obtained from 10 independent calculations.

\begin{figure*}
  \begin{minipage}{0.49\linewidth}
  \includegraphics[width=\textwidth]{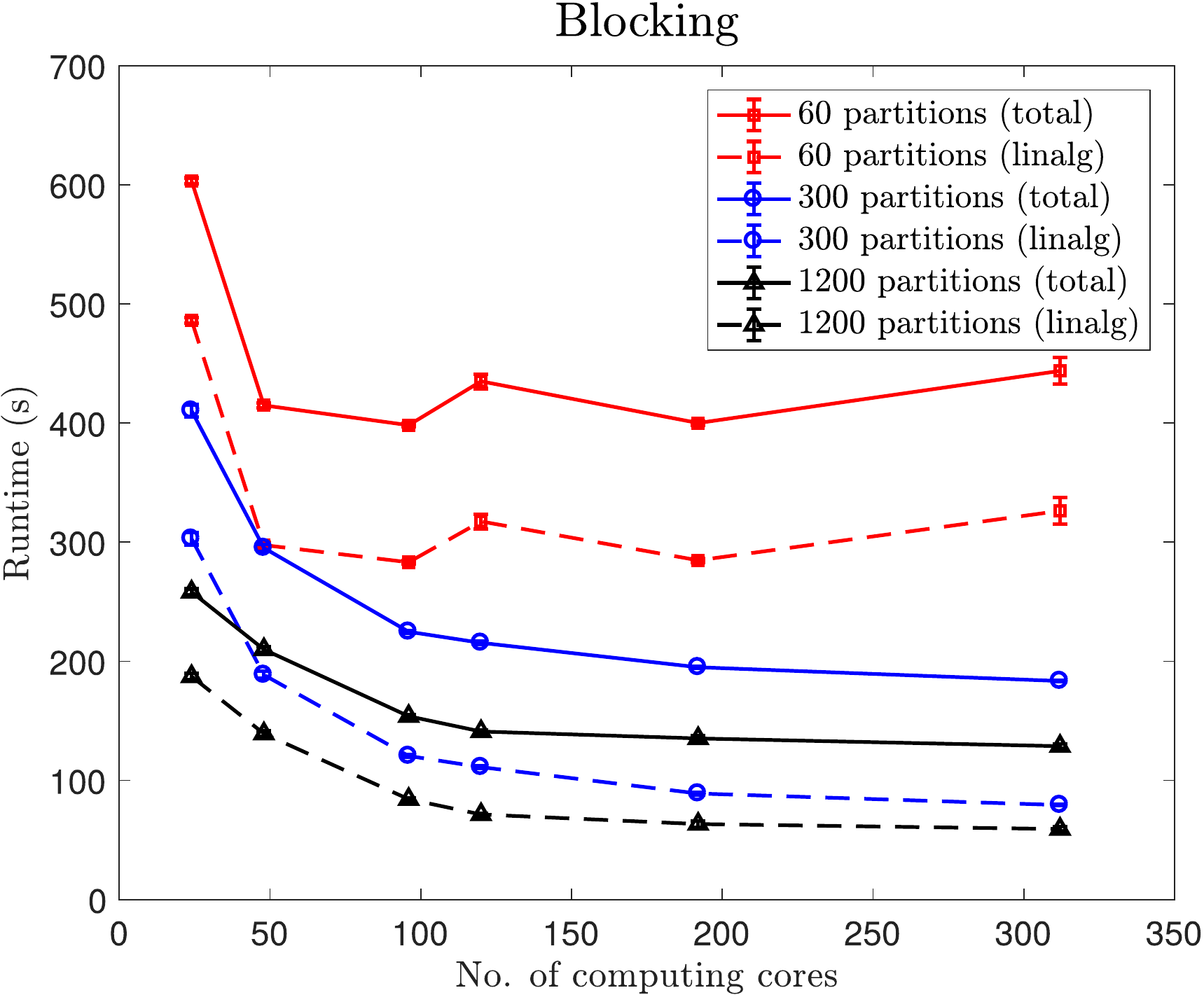}
  \subcaption{\label{fig:qmd_blocking}}
  \end{minipage}~
  \begin{minipage}{0.49\linewidth}
  \includegraphics[width=\textwidth]{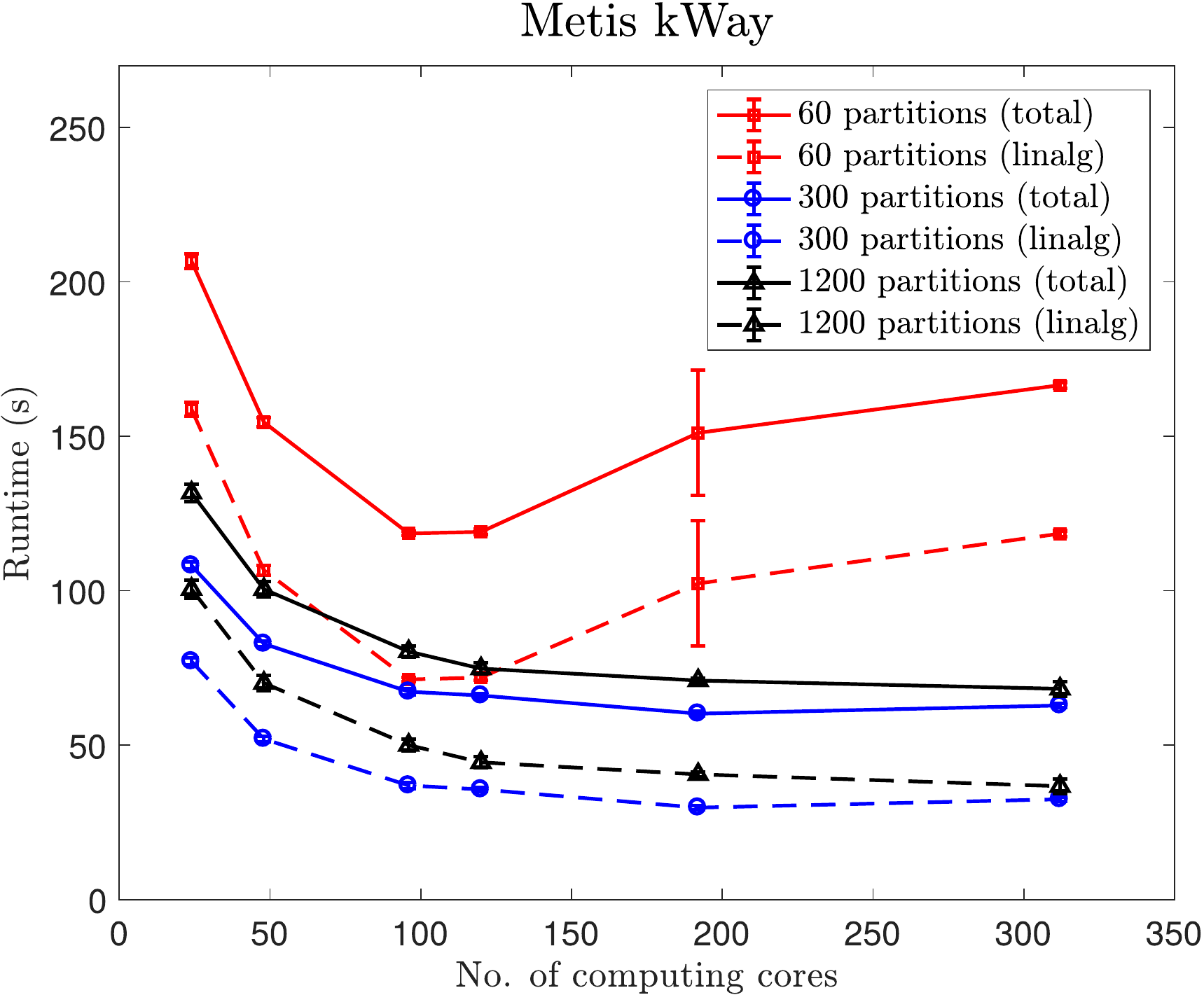}
  \subcaption{\label{fig:qmd_metis}}
  \end{minipage}
  \caption{Runtime for density matrix calculations of the \textit{polyalanine 289} protein test system as a function of the number of processors. Subproblems obtained from $60$, $300$ and $1200$ partitions computed using (a) the block partitioning approach and (b) \textit{METIS}. Runtimes for the total computation including communication costs are denoted by solid lines, and the times taken for linear algebra only are denoted by dotted lines.\label{fig:qmd_both}}
\end{figure*}

Fig.~\ref{fig:qmd_both} shows the total runtime taken to compute $\mathbf{P}$ for the \textit{polyalanine 289} protein system as the number of computing cores is varied. The graph was partitioned using a block partitioning approach (Fig.~\ref{fig:qmd_blocking}) as well as \textit{METIS} (Fig.~\ref{fig:qmd_metis}) into $60$, $300$ and $1200$ subproblems. First, a comparison of Figs.~\ref{fig:qmd_blocking} and \ref{fig:qmd_metis} demonstrates the advantage of using a tailored graph partitioning approach: Despite the additional effort for transforming the Hamiltonian into an undirected (and unweighted graph) and for partitioning it, using \textit{METIS} yields a substantial two-fold decrease (or more) in total runtime compared to the block partitioning approach. 

Figs.~\ref{fig:qmd_blocking} and \ref{fig:qmd_metis} also show the runtimes for independent SP2 subproblems if there were no overhead costs (i.e., the costs for linear algebra only). A significant fraction of the total runtime is indeed spent as overhead. A probable reason for the high overhead is that $\mathbf{H}$ and $\mathbf{P}$, as well as the associated connectivity graph, are initially processed serially so that all partitions were communicated from the \textit{mainChare} node to all others. Compared to sequential SP2, where a large distributed matrix is communicated after every SP2 iteration, the communication overhead for G-SP2 is smaller because matrices are communicated only at the beginning and at the end of SP2 iterations. However, this cost is still significant and it is consistent with our findings in the previous section where the shared memory \textit{CnC} application outperformed the \textit{Charm++} one.

\section{Conclusions}
\label{conclusions}
This article presents a way to improve the performance of electronic structure calculations that rely on accurate density matrix calculations. Traditionally, the density matrix $\mathbf{P}$ is obtained through methods that scale as $O(N^3)$, but  methods such as the second-order spectral projection (SP2) scale linearly. Very recently, a graph-based approach to SP2 was developed (G-SP2) to divide the computation of $\mathbf{P}$ into independent data-parallel subproblems executed on distributed-memory machines, but at the expense of causing load imbalances. We explored task-based implementations of the G-SP2 algorithm using \textit{CnC} and \textit{Charm++} asynchronous programming models. These implementations incorporate graph partitioning techniques developed in \cite{niklasson2016graph,GraphPaper} that are tailored for G-SP2. Task-based implementations of G-SP2 are important for mitigating load imbalances, especially if the same application is meant to be applied to different physical systems. 

We expect the approach presented in this paper to scale well with the number of atoms due to the nearsightedness of physical systems. The nearsightedness limits the maximum subproblem size as every core vertex in any partition can only have a limited number of halo vertices or neighbors. Therefore, as the size of the system increases, the number of partitions can be increased in order to limit the maximum subproblem size until an optimal number of partitions is reached. In our experiments we also observed a significant communication overhead which, nevertheless, can be mitigated by constructing submatrices of $\mathbf{H}$ locally, and by using a parallel version of the \textit{METIS} graph partitioning software. These are directions for future research.

We also investigate the effectiveness of different ways to generate subproblems. A simple locality based scheme can be advantageous in certain cases, but packages like \textit{METIS} are more general and take distance into account indirectly. Blocking is a na\"ive way to obtain partitions and can be inferior to locality based submatrices because partitions obtained by blocking contain core vertices that are arranged according to their index in the Hamiltonian and not according to any relevant physical quantity. However, if blocking is exchanged for a more effective technique like \textit{METIS}, the generation of subproblems becomes agnostic to how interactions between particular atoms vary with distance, and better subproblems (in terms of the computational cost metric used) are generated. Therefore, graph partitioning offers a general way to obtain subproblems based on the strength of interactions rather than distance between elements of $\mathbf{P}$. This feature is particularly useful for heterogenous systems with regions of dense and sparse connectivity. 

To conclude, we found our \textit{Charm++} based G-SP2 implementation to be more suitable for QMD simulations of large systems than the \textit{CnC} one as the former is readily extended to distributed-memory architectures. In addition to load balancing features, another advantage of using a mature runtime system such as \textit{Charm++} consists in the fact that further research in computer science to improve load balancing, fault tolerance or communication over large heterogenous clusters will directly carry over to our implementation without changes to the application code. Tuning \textit{Charm++} options to extract maximum performance from given hardware, expanding our tests to include large systems beyond 10,000 atoms, and a careful study of strong scaling limits for different physical systems could be future avenues for research.

\section*{Acknowledgments}
The authors would like to thank their mentors at Los Alamos National Laboratory for their help and feedback:
Ben Bergen,
Nick Bock,
Marc Cawkwell,
Hristo Djidjev,
Christoph Junghans,
Anders MN Niklasson,
and Ping Yang.

This work was supported by the Los Alamos Information Science \& Technology Center Institute (ISTI), the Advanced Simulation and Computing (ASC) Program and the US Department of Energy (DOE) and performed within the Co-Design Summer School 2015. Assigned: LA-UR-16-24908. Los Alamos National Laboratory, an affirmative action/equal opportunity employer, is operated by Los Alamos National Security, LLC, for the National Nuclear Security Administration of the US Department of Energy under contract DE-AC52-06NA25396.

This research used computing resources provided by the Los Alamos National Laboratory Institutional Computing Program, which is supported by the U.S.\ Department of Energy National Nuclear Security Administration.

Purnima Ghale is grateful for support and additional computing resources from the Center for Exascale Simulation of Plasma-Coupled Combustion (XPACC) at the University of Illinois, Urbana-Champaign, which is supported by the U.S.\ Department of Energy, National Nuclear Security Administration, Award Number DE-NA0002374.

\appendix
\section{Details on the Evaluation of Runtime Systems}
\label{long_task}

\subsection{Charm++}
\label{subsection_charm}
Here we present the salient features of \textit{Charm++}, followed by our \textit{Charm++} implementation of G-SP2. Additional details and a tutorial are available for \textit{Charm++} in~\cite{kalecharm++}.

\begin{enumerate}
  \item \textit{Charm++} is an object-oriented programming framework that consists of migratable objects called \textit{chares}.  These objects are \textit{C++} objects, contain data and operations, of which some can be invoked from other \textit{chares} via proxies. A \textit{Charm++} program starts with a \textit{main chare} that interacts with the user, files, or other programs to obtain input data and instructions. The \textit{main chare} then produces other \textit{chare} workers that execute program tasks. \textit{Chares} of the same type can be arranged in a collection called a \textit{chare array}.

  \item \textit{Charm++} tolerates latency by executing worker \textit{chares} only when an execution call is received from another object and supports fault tolerance by automating checkpoints and restarts. Operations prescribed in a \textit{chare} object are executed only when an execution call is received from another object (as in object-oriented programming) and when all data dependencies are satisfied. This is a useful mechanism for tolerating latency as a process is allocated to a processor only when a message for its execution is received. When using \textit{chare arrays}, the same message can be broadcast to all the elements in the \textit{chare array} at once. The runtime system then dynamically schedules and balances the processes to be run.

  \item Both shared- and distributed-memory architectures can be targeted. Typically desired functionalities such as load balancing, memory management, detection of program initialization and termination, and hardware interfacing are available. The programmer can either choose to tune using various load-balancing schemes (or other functionalities) provided by \textit{Charm++} or can provide a self-defined scheme. 
\end{enumerate}

\begin{figure}[t]
\centering
\includegraphics[width=0.6\columnwidth]{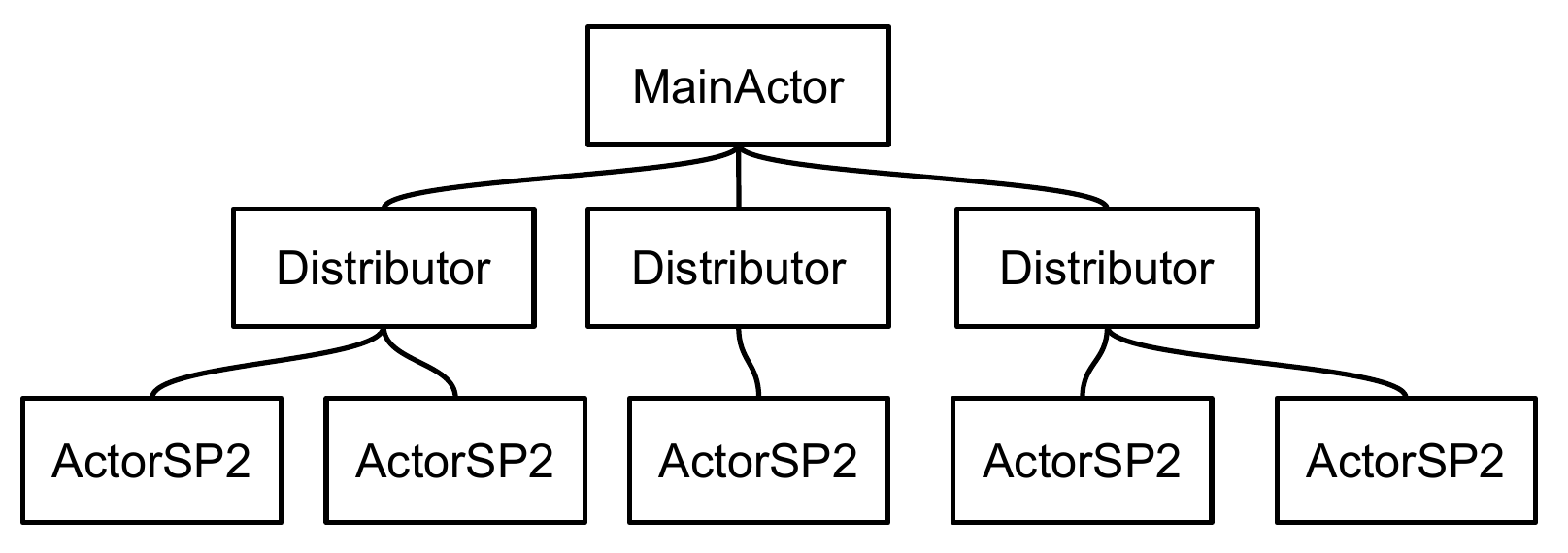}
\caption{\label{fig:charmpp}Roles of different {chare} objects constructed in \textit{Charm++} for the target G-SP2 application described in the main text.}
\end{figure}

Our \textit{Charm++} program uses three types of objects which we call \textit{MainActor}, \textit{Distributors} and \textit{ActorSP2}. As shown in Fig.~\ref{fig:charmpp}, the \textit{main chare} called \textit{MainActor} reads the Hamiltonian and other input parameters. It then sends the collected data to distributors that create their own \textit{chare arrays} that receive and execute one SP2 subproblem. Computing nodes were assigned to distributors by round-robin scheme. Results from the individual SP2 subproblems are then sent to the \textit{MainActor} which assembles the full density matrix $\mathbf{P}$. 

For load balancing, we used the \textit{GreedyCommLB} scheme of \textit{Charm++}. \textit{Charm++} also includes \textit{MPI-interoperate} functions, which can compile the above implementation as an ordinary library that can be called from standard \textit{MPI} code. Our implementation of \textit{Charm++} also provides the option to compile the entire program as an \textit{MPI} library that can been called from any \textit{MPI} program using the advanced \textit{MPI-interoperate} functions in \textit{Charm++}. 

\subsection{Concurrent Collections CnC}
With \textit{Charm++}, an application is decomposed into migratable \textit{C++} objects that interact with each other through method calls (see section \ref{subsection_charm}). In contrast, \textit{CnC} (\textit{Concurrent Collections}) decomposes the application logic into a flow of data and control commands which, during execution, invoke various computations. Thus, \textit{CnC} allows the programmer to focus on expressing the program at a higher level, while giving flexibility to the runtime system to schedule specific operations. We used the Intel \textit{CnC} runtime system, which provides the programmer with a variety of scheduling options. The scheduler can also interact with the environment via flags or environment variables. The Intel \textit{CnC} runtime system allows thread-sharing and may also spawn internal helper threads. 

The salient features of the \textit{CnC} programming model are summarized below. More detailed information can be found in \cite{burke2011concurrent,Budimlic2010}.

\begin{enumerate}
  \item The dependency graph, also known as the \textit{CnC specification graph} expresses the application logic with the help of objects and their tags: (a) \textit{step collections}, which are operations to be executed; (b) \textit{data collections}, which determine data flows; and (c) \textit{control collections}, which act as switches that determine the order in which computations and data flow are allowed. Standard convention for representing these three different object types and their dependencies in the specification graph is with ellipses, rectangles, and hexagons, respectively. Collections are tracked by their respective tags.

  \item The \textit{CnC} philosophy holds that computations with dependencies can be scheduled concurrently. This is implemented by identifying computations that either produce or consume certain data, or ones that control or are controlled by another computation. While such an approach frees a wide range of possibilities to exploit parallelism, it also increases the complexity of the application since data, controls, and computations exist independently. As an aid to the programmer, templates for application-specific tuners are available, where the tuning expert can add additional soft constraints to guide the runtime.
 
  \item Control and data flows must be explicitly defined in a \textit{CnC} program via the API (the Application Program Interface which specifies how software components should interact with each other, typically through a set of routines and protocols). Because \textit{CnC} assumes distributed-shared memory (i.e., it allows global variables), modifications to a shared data object must follow a control logic, making \textit{CnC} deterministic \cite{Budimlic2010}. 
\end{enumerate}

\begin{figure*}
\centering
\includegraphics[width=\textwidth]{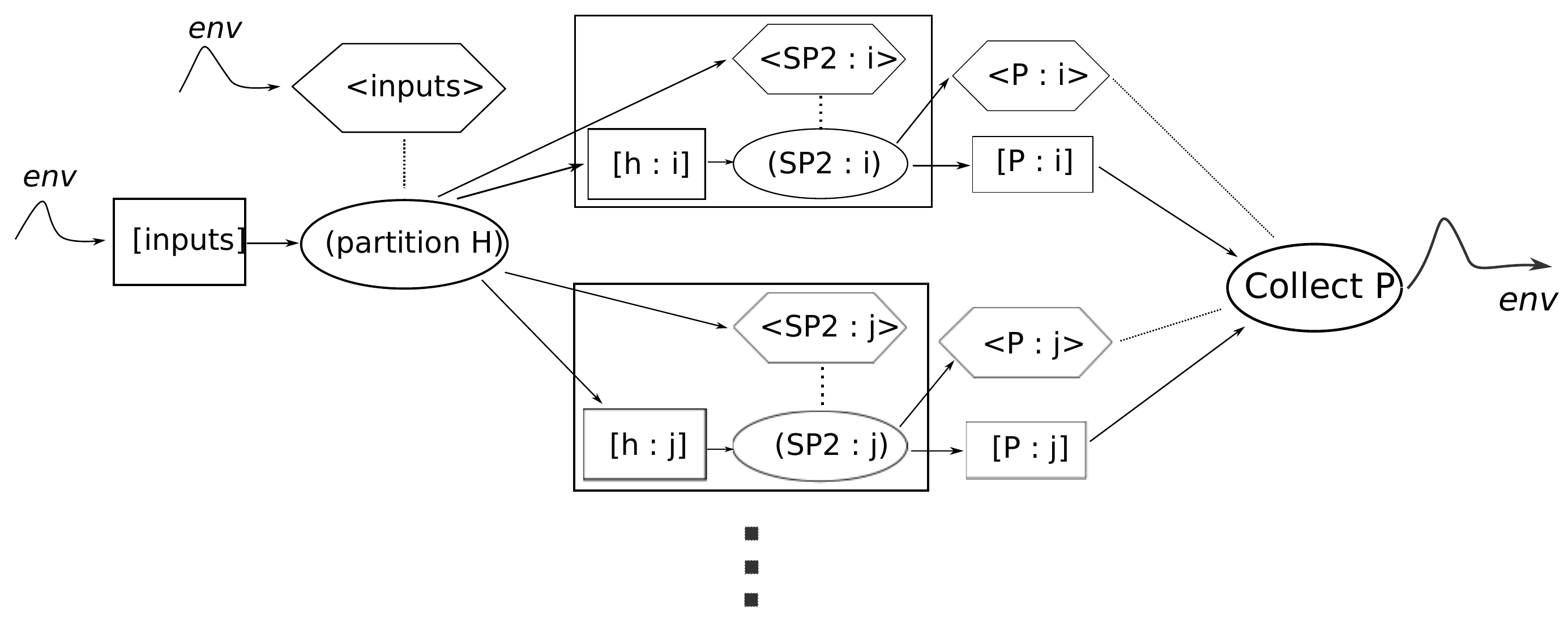}
\caption{The specification graph for our \textit{CnC} implementation of G-SP2.\label{fig:CnCspec}}
\end{figure*}

In our G-SP2 implementation, we optimized the runtime for execution on a shared-memory machine. Thus, we accessed global variables and pointers within the steps, which in turn made the porting to a distributed-memory implementation more complicated. 
Our \textit{CnC} specification graph is shown in Fig.~\ref{fig:CnCspec}. The environment provides input data as well as the control tag to initiate the program. The input data consists of both the Hamiltonian matrix and information to divide it into subproblems. The given Hamiltonian is then split into subproblems, which generates many instances of SP2 computations. Each SP2 computation, labeled by $i$, requires a partition of the Hamiltonian $[\text{h : i}]$, and a control prescription $<\text{SP2 : i}>$ and produces partitions of the density matrix $[\text{P : i}]$, which are then collected, and passed to the environment. During execution, input parameters are provided by the user or the environment. Each unique input set gives rise to an instance of the application, which is dynamically load balanced by the runtime system.

% \bibliographystyle{apalike}
% \bibliography{combined}

\end{document}